\documentclass{article}
\usepackage{psfig}
\usepackage{amsmath}
\usepackage{ifthen}

\newcommand{\eq}[2][]{
 \ifthenelse{\equal{#1}{}}{
 \begin{equation*}
 #2
 \end{equation*}
 }{
 \begin{equation}
 \label{#1}
 #2 
 \end{equation}
 }
}

\newcommand{\meq}[2][{}]{
 \ifthenelse{\equal{#1}{}}{
 \begin{equation*}
 \begin{split}
 #2
 \end{split}
 \end{equation*}
 }{
 \begin{equation}
 \label{#1}
 \begin{split}
 #2
 \end{split}
 \end{equation}
 }
}

\newcommand{\de}{\partial}
\newcommand{\eps}{\varepsilon}

\newcommand{\AND}{\wedge}
\newcommand{\OR}{\vee}
\newcommand{\XOR}{\oplus}

\renewcommand{\vec}[1]{\boldsymbol{#1}}

\begin{document}

\title{Nature of phase transitions in a probabilistic cellular
 automaton with two absorbing states}

\author{Franco Bagnoli \\
 Dipartimento di Matematica Applicata, \\
 Universit\`a di Firenze, Via S. Marta 3, I-50139 Firenze,
 Italy \\
 and INFN and INFM, Sez. di Firenze.\\
 bagnoli@dma.unifi.it\\
\\
Nino Boccara\\
 DRECAM/SPEC, CE-Saclay, F-91191 Gif-sur-Yvette Cedex,
 France \\ 
 and Depart\-ment of Physics, University of Illinois, Chicago,
 USA \\
 nboccara@amoco.saclay.cea.fr, boccara@uic.edu\\
\\
Ra\'ul Rechtman \\
 Centro de Investigac\'{\i}on en Energ\'{\i}a, UNAM, \\
 62580 Temixco, Morelos, Mexico \\
 rrs@teotleco.cie.unam.mx
}

\date{\today}
\maketitle

\begin{abstract} 
We present a probabilistic cellular automaton with two absorbing
states, which can be considered a natural extension of the
Domany-Kinzel model. Despite its simplicity, it shows a very rich
phase diagram, with two second-order and one first-order transition
lines that meet at a bicritical point. We study the phase
transitions and the critical behavior of the model using mean field
approximations, direct numerical simulations and field theory.
 The second-order critical curves and
the kink critical dynamics are found to be in the directed
percolation and parity conservation universality classes,
respectively. The first order phase transition is put in evidence 
by examining the hysteresis cycle. We also study the ``chaotic''
phase, in which two replicas evolving with the same noise diverge,
using mean field and numerical techniques. Finally, we show how the
shape of the potential of the field-theoretic formulation of the
problem can be obtained by direct numerical simulations.
\end{abstract}

\section{Introduction}
Probabilistic cellular automata (PCA) have been widely used to model
a variety of systems with local interactions in physics, chemistry,
biology and social
sciences~\cite{Farmer_etal_1984,Wolfram_1986,Manneville_etal_1989,%
Gutowitz_1990,Boccara_etal_1993}. Moreover, PCA are simple and
interesting models that can be used to investigate fundamental
problems in statistical mechanics. Many classical equilibrium spin
models can be reformulated as PCA, for example the kinetic Ising
model with parallel heat-bath dynamics is strictly equivalent to a
PCA with local parallel
dynamics~\cite{Bagnoli_etal_1991,Bagnoli_1996}. On the other hand,
PCA can be mapped to spin models~\cite{Georges_1989} by expressing
the transition probabilities as exponentials of a local energy. PCA
can be used to investigate nonequilibrium phenomena, and in
particular the problem of phase transitions in the presence of
absorbing states. An absorbing state is represented by a set of
configurations from which the system cannot escape, equivalent to an
infinite energy well in the language of statistical mechanics. A
global absorbing state can be originated by one or more local
transition probabilities which take the value zero or one, 
corresponding to some infinite coupling in the local
energy~\cite{Georges_1989}. 

The Domany-Kinzel (DK) model is a boolean PCA on a tilted square 
lattice that has been extensively 
studied~\cite{Domany_Kinkel_1984,Kinzel_1985}. Let us denote the two 
possible states of each site with the terms ``empty'' and
``occupied''. In this model a site at time $t$ is connected to two
sites at time $t-1$, constituting its neighborhood. The control
parameters of the model are the local transition probabilities that
give the probability of having an occupied site at a certain
position once given the state of its neighborhood. The transition
probabilities are symmetric for all permutations of the neighborhood,
and this property is equivalent to saying that they depend on the
sum of ``occupied'' sites in the neighborhood, whence the term
``totalistic'' used to denote this class of automata.

In the DK model the transition probability from an empty neighborhood
to an occupied state is zero, thus the empty configuration is an
absorbing state. For small values of the other transition
probabilities, any initial configuration will evolve to the absorbing
state. For larger values, a phase transition to an active phase,
represented by an ensemble of partially occupied configurations, is
found. The order parameter of this transition is the asymptotic
average fraction of occupied sites, which we call the density. The
critical properties of this phase transition belong to the directed
percolation (DP) universality class (except for one extreme
point)~\cite{DP}, and the DK model is often considered the prototype
of such a class. 

The evolution of this kind of models is the discrete equivalent of
the trajectory of a stochastic dynamical system. One can determine
the sensitivity with respect to a perturbation, by studying the
trajectories originating by two initially different configurations
(replicas) evolving with the same realization of the stochasticity,
e.g.\ using the same sequence of random numbers. The order parameter
here is the asymptotic difference between the two replicas, which we
call the damage. It turns out that, inside the active phase, there
is a ``chaotic'' phase in which the trajectories depend on the
initial configurations and the damage is different from zero, and a
non chaotic one in which all trajectories eventually synchronize with
the vanishing of the damage. In simple models like the DK one, this
transition does not depend on the choice of the initial
configurations (provided they are different from the absorbing state)
and the initial damage~\cite{Martins_etal_1991}.

It has been conjectured that all second-order phase transitions from
an ``active'' phase to a non degenerate, quiescent phase (generally
represented by an absorbing state) belong to the DP universality
class if the order parameter is a scalar and there are no extra
symmetries or conservation
laws~\cite{Janssen_1981,Grassberger_1982}. This has been verified in
a wide class of models, even multi-component, and in the presence of
several asymmetric absorbing states~\cite{_Hinrichsen_1997}. Also
the damage phase transition has a similar structure. Once
synchronized, the two replicas cannot separate, and thus the
synchronized state is absorbing. Indeed, numerical simulations shows
that it is in the DP universality class~\cite{Grassberger_1985}.
Moreover, in the DK model, the damage phase transition can be
mapped onto the density one~\cite{Bagnoli_1996}. 

On the other hand, some models with conserved
quantities~\cite{Grassberger_etal_1984,Grassberger_1989} or symmetric
absorbing states belong to a different universality class called
parity conservation (PC) or directed
Ising~\cite{Hinrichsen_1997,Hwang_etal_1998}. This universality class
appears to be less robust since it is strictly related to the
symmetry of the absorbing states; a slight asymmetry is sufficient to
bring the model to the usual DP
class~\cite{Hinrichsen_1997,Hwang_etal_1998}.

An interesting question concerns the simplest, one-dimensional PCA
model with short range interactions exhibiting a first order phase
transition. Dickman and
Tom\'e~\cite{Dickman_Tome_1991,Dickman_Marro_1999} proposed a contact
process with spontaneous annihilation, auto catalytic creation by
trimers and hopping. They found a first order transition for high
hopping probability, i.e.,\ in the region more similar to mean field 
(weaker spatial correlations).

Bassler and Browne discussed a model 
whose phase diagram also presents first and second-order
phase transitions~\cite{Bassler_Browne_1996}.
In it, monomers of three
different chemical species can be adsorbed on a one-dimensional surface
and neighboring monomers belonging to different species annihilate
instantaneously. The control parameters of the model are the absorption
rates of the monomers. The transition from a saturate to a reactive
phase belongs to the DP universality class, while the transition
between two saturated phases is discontinuous. The point at which
three phase transition lines join does belong to the PC universality
class. 

Scaling and fluctuations near first order phase transitions are also
an interesting subject of 
study~\cite{Ehsasi_1989,Evans,Monetti_etal_1999,Hinrichsen_etal_1999},
which can profit from the existence of simple models.

In this paper we study a one-dimensional, one-component, totalistic
PCA with two absorbing states. It can be considered as a natural
extension of the DK model to a lattice in which the neighborhood of
a site at time $t$ contains the site itself and its two nearest
neighbors at time $t-1$. This space-time lattice arises naturally in
the discretization of one-dimensional reaction-diffusion systems. In
our model, the transition probabilities from an empty neighborhood is
zero, and that from a completely occupied neighborhood is one. The
model has two absorbing states: the completely empty and the
completely occupied configurations. The order parameter is again the
density; it is zero or one in the two quiescent phases, and assumes
other values in the active phase. The system presents a line of
symmetry in the phase diagram, over which the two absorbing phases
have the same importance. A more detailed illustration of the model
can be found in Section~\ref{sec:model}.

This model can arise as a particular case of a nonequilibrium
wetting of a surface. In this framework, only a single layer of 
particles can be absorbed on the surface. If we assume that particles
can be absorbed or desorbed only near the boundaries of a patch of
already absorbed particles (when the neighborhood is not
homogeneous), then the completely empty and occupied configurations
are absorbing states. 

This totalistic PCA can also be interpreted as a simple model of
opinion formation. It assumes that an individual may change his mind
according to himself and his two nearest neighbors. The role of
social pressure is twofold. If there is homogeneity of opinions,
individuals cannot disagree (absorbing states), otherwise they can
agree or disagree with the majority with a certain probability.

The density phase diagram shows two second order phase transition
curves separating the quiescent phases from the active one, and a
first order transition line between the two quiescent phases, as 
discussed in Section~\ref{sec:phasediagram}. These curves meet on
the line of symmetry in a bicritical point. We use both mean field
approximations and direct numerical simulations. The former simple
approximation gives a qualitatively correct phase diagram. The
numerical experiments are partially based on the fragment
method~\cite{Bagnoli_etal_1997}. This is a parallel algorithm that
implements directly the evolution rule for different values of the
control parameters on the bits of one or more computer words.

In Section~\ref{sec:critical}, we investigate numerically the second
order phase transitions and find they belong to the DP
universality class. Along the line of symmetry of the model the two
absorbing phases are equivalent. In Appendix~\ref{app:kinks} we
show that on this line one can reformulate the problem in terms of
the dynamics of kinks between patches of empty and occupied sites. 
Since the kinks are created and annihilated in pairs, the dynamics
conserves the initial number of kinks modulo two. In this way we can
present an exact mapping between a model with symmetric absorbing
phases and one with parity conservation. We find that the critical
kink dynamics at the bicritical point belongs to the PC
universality class. 

In Section~\ref{sec:chaotic} we study the chaotic phase, using
dynamic mean field techniques (reported in 
Appendix~~\ref{app:meandamage}) and direct numerical simulations. The
location of this phase is similar to that of the DK model: it joins
the second-order critical curves at the boundary of the phase
diagram. 

Our model exhibits a first-order phase transition along the line of
symmetry in the upper part of the phase diagram. A first-order
transition is usually associated to an hysteresis cycle. It is
possible to observe such a phenomena by adding a small perturbing
field to the absorbing states, as discussed in
Section~\ref{sec:firstorder}.

The DP universality class is equivalent to the Reggeon field
theory~\cite{Cardy_Sugar_1980}, which in $d=0$ corresponds to a
quadratic potential with a logarithmic divergence at the origin. The
Langevin description for systems in the PC class yields a similar
potential, except for irrelevant terms~\cite{Grassberger_1989}. It
has been shown~\cite{Munoz_1998} that one can reconstruct the
potential from the numerical integration of the Langevin equation,
which, however, requires special techniques in the presence of
absorbing states~\cite{Dickman_1994}. In
Section~\ref{sec:reconstruction} we show how the potential is
reconstructed from actual simulations of a phenomenological model,
such as our original cellular automata or the kink dynamics. In this
way we obtain the shape of the potential for a system in the parity
conservation universality class.

\section{The model}\label{sec:model}
We describe here a one-dimensional, totalistic, probabilistic
cellular automaton with three inputs. The state of the model at time
$t$ is given by $\boldsymbol{x}^t=(x_0^t,\dots,x_{L-1}^t)$ with
$x_i^t\in \{0,1\}$; $t=1,2,\dots$ and $L$ is the number of sites. All
operations on spatial indices are assumed to be modulo $L$ (periodic
boundary conditions). For simplicity of notation, we write
$x=x_i^t$, $x_- = x_{i-1}^t$, $x_+ = x_{i+1}^t$ and $x'=x_i^{t+1}$. 
We shall indicate by $\sigma = x_- + x + x_+$ number of occupied cells in
the neighborhood.
The most general three-input totalistic PCA is defined by the
quantities $p_s$, which are the conditional 
probabilities that $x'=1$ if $\sigma=s$. The microscopic dynamics of
the model is completely specified by
\eq[x]{
 x' = \sum_{s=0}^3 R_s \delta_{\sigma,s}.
}
In this expression $R_s$ is a stochastic binary variable that takes
the value 1 with probability $p_s$ and 0 with probability $1-p_s$,
and $\delta$ is the Kronecker delta. In practice, we implement $R_s$
by extracting a random number $r_s$ uniformly distributed between 0
and 1, and setting $R_s$ equal to 1 if $r_s<p_s$ and 0 otherwise.
Eq.~\eqref{x} implies the use of four random numbers $r_s$ for each
site. The evolution of a single trajectory is not affected by the
eventual correlations among the $r_s$, since only one 
$\delta_{\sigma,s}$ is different from zero. This is not true when
computing the simultaneous evolution of two or more replicas using
the same noise. If not otherwise stated, we use only one random
number for all the $r_s$. More discussions about the choice of
random numbers can be found in Section~\ref{sec:chaotic} and in
Appendix~\ref{app:kinks}.

With $p_0=0$ and $p_3=1$ the model presents two quiescent phases: the
phase 0 corresponding to the configuration $x=(0,\dots, 0)$ and the
phase 1 corresponding to the configuration $x=(1,\dots, 1)$. In
this case there are two control parameters, $p_1$ and $p_2$, and the
model is symmetric under the changes $p_1\to 1-p_2$, $p_2\to 1-p_1$
and $x\to x\XOR 1$ where $\XOR$ is the exclusive disjunction (or the
sum modulo 2).

\section{Phase diagram}\label{sec:phasediagram}

\begin{figure}[t]
 \centerline{\psfig{figure=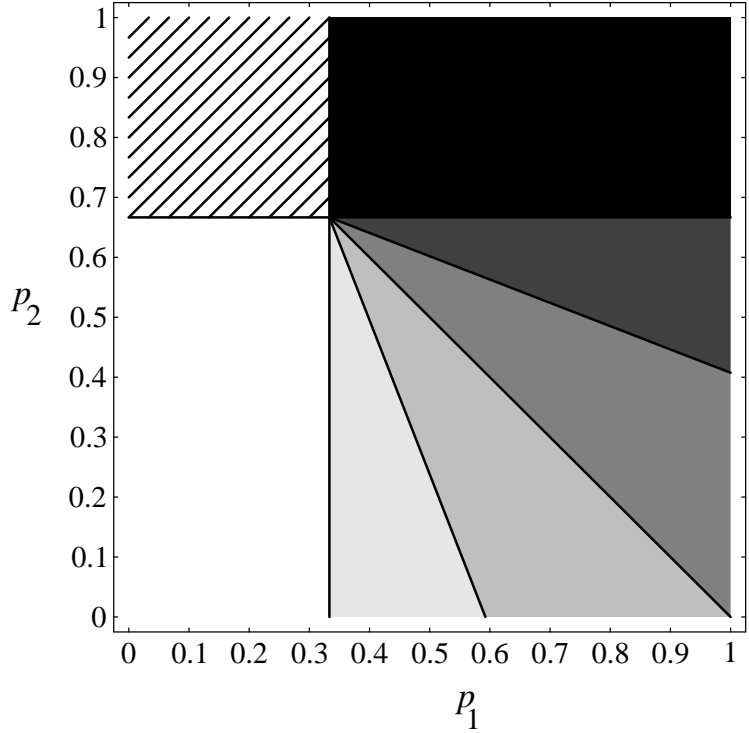,width=10cm}}
 \caption{Mean-field phase diagram for the density 
 $c$ of active sites. The white (black) region corresponds to the phase 0
 (phase 1). The levels of grey indicate different values of the
 asymptotic density $c$ (active phase),
 the lightest corresponds to $0<c<1/4$, and the
 next ones to $1/4<c<1/2$, $1/2<c<3/4$ and $3/4<c<1$. The two 
 quiescent phases coexist in the hatched region.}
 \label{fig:meanfield} 
\end{figure}

In order to have a qualitative idea of the behavior of the model, we first 
study the mean-field approximation. If $c$ and $c'$ denote the density of
occupied sites at times $t$ and $t+1$ respectively,
\meq[eq:c]{ 
 c'&=3p_1c\left(1-c\right)^2+3p_2c^2\left(1-c\right)+c^3.
}
This map has three fixed points, $c_0$, $c_1$, and $c_2$ given by
\eq{
 c_0=0,\quad c_1=\frac{3p_1-1}{1+3p_1-3p_2},\quad\hbox{and}\quad c_2=1.
}
The asymptotic density will assume one of the latter values according
to the values of the control parameters and the initial state as we
show in Fig.~\ref{fig:meanfield}. In the square $1/3<p_1\leq 1$,
$0\leq p_2<2/3$, the only stable fixed point is $c_1$. Inside this
square, on the segments $p_2-2/3=m(p_1-1/3)$ with $m<0$,
$c_1=1/(1-m)$.The first fixed point $c_0$ is stable when $p_1<1/3$
and $c_2$ is stable when $p_2>2/3$. There is a continuous
second-order transition from the quiescent phase 0 to the active
phase on the segment $p_1=1/3$, $0\leq p_2 < 2/3$ and another
continuous transition from the active to the quiescent phase 1 on
the segment $1/3< p_1\leq 1,p_2=2/3$.

In the hatched region of Fig.~\ref{fig:meanfield} $c_0$ and $c_2$ are
both stable. Their basins of attraction are, respectively, the
semi-open intervals $[0,c_1)$ and $(c_1,1]$. Starting from a
uniformly distributed random value of $c$, as time $t$ goes to
infinity, $c$ tends to $c_0$ with probability $c_1$, and to $c_2$
with probability $1-c_1$. Since, for $p_1+p_2=1$, $c_1=1/2$, the
segment $p_1+p_2=1$, with $0\le p_1<1/3$ and $2/3<p_2\le 1$, is
similar to a first-order transition line between the phase 0 and the
phase 1.

\begin{figure}[t]
 \centerline{\psfig{figure=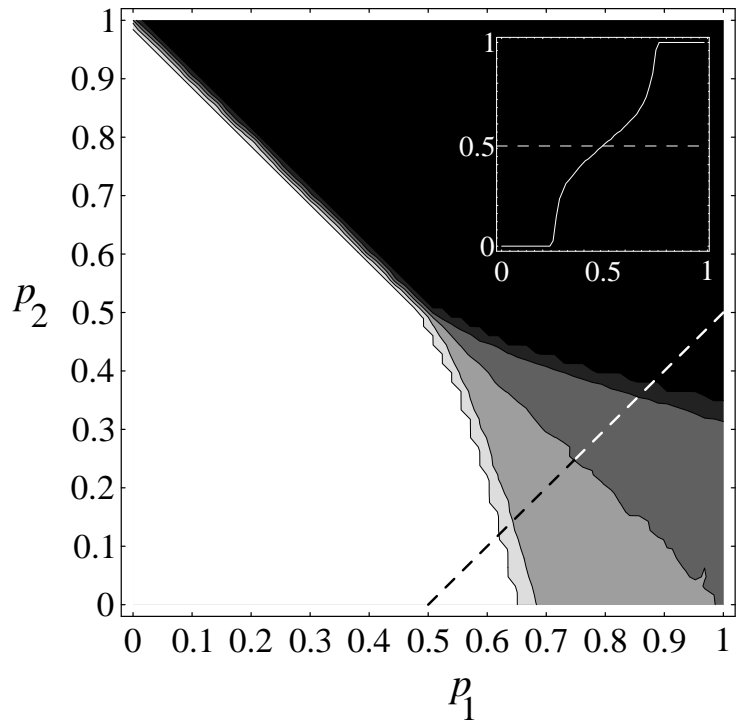,width=10cm}}
 \caption{
 Phase diagram for the density of active sites $c$ by
 numerical experiments. One run was performed with $L=10000$ and
 $T=10000$. The graph shows $64\times 64$ values of $p_1$ and
 $p_2$. The color code is the same as in
 Fig.~\protect{\ref{fig:meanfield}}. The inset represents the 
 density profile along the dashed line. Two critical phase transitions 
 are evident.
 }
 \label{fig:phasediagram}
\end{figure}

In Fig.~\ref{fig:phasediagram} we show the phase diagram of the
model obtained numerically starting from a random initial state with
half of the cells occupied. The scenario is qualitatively the same
as predicted by the mean-field analysis. In the vicinity of the point
$(p_1,p_2)=(0,1)$ we observe a discontinuous transition from $c=0$ to
$c=1$. The two second-order phase-transition curves from the active
phase to the quiescent phases are symmetric, and the critical
behavior of the order parameter, $c$ for the lower curve and $1-c$ 
for the upper one, is the same. 

Due to the symmetry of the model the two second-order phase
transition curves meet at a bicritical point $(p_t,1-p_t)$ where
the first-order phase transition line $p_1+p_2=1$, $p_1<p_t$ ends.
Crossing the second-order phase boundaries on a line parallel to the
diagonal $p_1=p_2$, the density $c$ exhibits two critical
transitions, as shown in the inset of Fig.~\ref{fig:phasediagram}.
Approaching the bicritical point the critical region becomes
smaller, and corrections to scaling increase. Finally, at the
transition point the two transitions coalesce into a single
discontinuous one. 

\section{Critical dynamics and universality
classes}\label{sec:critical}

We performed standard dynamic Monte Carlo simulations starting from a
single site in the origin out of the nearest absorbing state, and 
measured the average number of active sites $N(t)$, the survival
probability $P(t)$ and the average square distance from origin
$R^2(t)$ (averaged over surviving runs) defined as
\meq[eq:crit]{ 
 N(t) &= \dfrac{1}{K}\sum_{k=1}^K \sum_{i} \omega_i^t(k) , \\
 P(t) &= \dfrac{1}{K}\sum_{k=1}^K \theta\left(\sum_{i}
 \omega_i^t(k)\right),\\
 R^2(t) &= \dfrac{1}{KN(t)} \sum_{k=1}^K \sum_{i}
 \omega_i^t(k) i^2.
}
In these expressions, $k$ labels the different runs and $K$ is the
total number of runs. The quantity $\omega_i$ is $x_i$ if the
nearest absorbing state is $\vec{0}$ and $1-x_i$ otherwise; $\theta$
is the Heaviside step function, that assumes the value 1 if its
argument is greater than 0, and the value 0 if it is smaller than 0.

At the critical point one has
\eq{
 N(t) \sim t^\eta,\quad P(t) \sim t^{-\delta},\quad R^2(t) \sim t^z.
}
At the transition point 
$(p^*_1=0.6625(3),p_2=0)$, we get $\eta=0.308(5)$, $\delta=0.160(2)$ and
$z=1.265(5)$, in agreement with the best known values
for the directed percolation universality class~\cite{Munoz_etal_1999}. 

Near the bicritical point, on the line $p_1+p_2=1$, the two
absorbing states have symmetrical weight. We define a kink $y_i$
as $y_i=x_i\XOR x_{i+1}$. For the computation of the critical
properties of the kink dynamics, one has to replace $\omega_i$ with
$y_i$ in Eq.~\eqref{eq:crit}. The evolution equation is derived in
Appendix~\ref{app:kinks}. In the kink dynamics there is only one
absorbing state (the empty state), corresponding to one of the two
absorbing states $\vec{0}$ or $\vec{1}$. For $p_1<p_t$ the
asymptotic value of the density of kinks is zero and it starts to
grow for $p_1>p_t$. In models with multiple absorbing states,
dynamical exponents may vary with initial conditions. Quantities
computed only on survival runs ($R^2(t)$) appear to be universal, 
while others (namely $P(t)$ and $N(t)$) are not~\cite{Jensen_1994}. 

We performed dynamic Monte Carlo simulations starting either from
one and two kinks. In both cases $p_t=0.460(2)$, but the exponents 
were found to be different. Due to the conservation of the number
of kinks modulo two, starting from a single site one cannot observe
the relaxation to the absorbing state, and thus $\delta=0$. In this
case $\eta =0.292(5)$, $z=1.153(5)$. On the other hand, starting
with two neighboring kinks, we find $\eta=0.00(2)$,
$\delta=0.285(5)$, and $z=1.18(2)$. These results are consistent with
those found by other 
authors~\cite{Grassberger_etal_1984,Grassberger_1989,Hinrichsen_1997}.

\section{The chaotic phase}\label{sec:chaotic}

\begin{figure}[t]
 \centerline{\psfig{figure=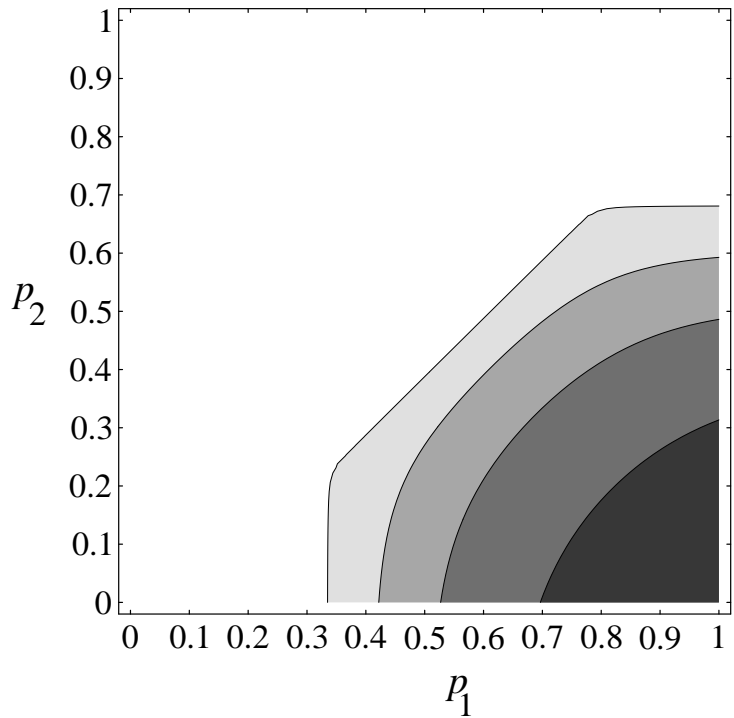,width=10cm}}
 \caption{Mean-field damage-spreading phase diagram. The diagram has been 
 obtained numerically iterating Eq.~\protect{\eqref{eq:eta}}. The
 lightest level of grey corresponds to $0<h<1/8$, the next ones to
 $1/8<h<1/4$, $1/4<h<3/8$ and $3/8<h<1/2$ respectively.
 }
 \label{fig:meandamage}
\end{figure}

Let us now turn to the sensitivity of the model to a variation in the
initial configuration, i.e.\ to the study of damage spreading or,
equivalently, to the location of the chaotic phase.

Given two replicas $x$ and $y$, we define the difference $w$ as
$w=x\XOR y$. The damage $h$ is defined as the fraction of sites in
which $w=1$, i.e.\ as the Hamming distance between the configurations
$x$ and $y$.

The precise location of this phase transition depends on the
particular implementation of the stochasticity. Since the sum of
occupied cells in the neighborhood of $x$ is in general different
from that of $y$, the evolution equation \eqref{x} for the two
replicas uses different random numbers $r_s$. The correlations among
these random numbers affect the location of the chaotic phase
boundary~\cite{Domany_Hinrichsen}. 

We limit our investigation to the case of maximal correlations by
using just one random number per site, i.e.\ all $r_s$ are the same
for all $s$ at the same site. This gives the smallest possible
chaotic region. Notice that we have to extract a random number for
all sites, even if in one or both replicas have a neighborhood
configuration for which the evolution rule is deterministic (if
$\sigma=0$ or $\sigma=3$). 

One can write the mean field equation for the damage by taking into
account all the possible local configurations of two lattices. The
evolution equation for the damage depends on the correlations among
sites but in the simplest case we can assume that $h(t+1)$ depends
only on $h(t)$ and $c(t)$, the density of occupied sites. In
Appendix~\ref{app:meandamage} we find the evolution equation for the
damage in the mean-field approximation. In 
Fig.~\ref{fig:meandamage} we show the phase diagram of the chaotic
phase in this approximation. There is a qualitative agreement with
the mean field phase diagram found for the DK
model~\cite{Tome_1994}.

\begin{figure}[t]
 \centerline{\psfig{figure=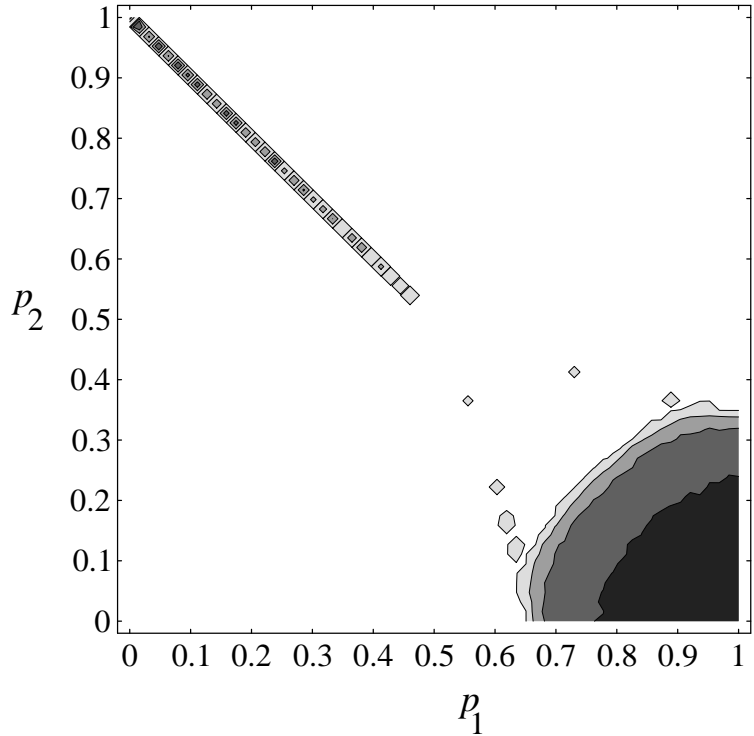,width=10cm}}
 \caption{Phase diagram for the damage spreading 
 from direct numerical simulations. The color code is that of 
 Fig.~\ref{fig:meandamage} Traces of the second order phase transitions
 are present; they join to the first-order ($c_0=0.5$) phase boundary. 
 }
 \label{fig:damage}
\end{figure}

In Fig.~\ref{fig:damage} we show the phase diagram for the damage
found numerically by considering the evolution starting from
uncorrelated configurations with initial density equal to 0.5. The
damage region is shown in shades of grey. Outside this region there
appear small damaged domains on the other phase boundaries. This is
due either to the divergence of the relaxation time (second-order
transitions) or to the fact that a small difference in the initial
configuration can drive the system to a different absorbing state
(first-order transitions). The chaotic domain near the point
$(p_1,p_2)=(1,0)$ is stable regardless of the initial density. On
the line $p_2=0$ the critical points of the density and the damage 
coincide at $p^*_1$.

\section{First-order phase transition and hysteresis cycle}\label{sec:firstorder}

First-order phase transitions are usually associated to a hysteresis
cycle due to the coexistence of two phases. In the absence of
absorbing states, the coexistence of two stable phases for the same
values of the parameters is a transient effect in finite systems,
due to the presence of fluctuations. To find the hysteresis loop we
modify the model slightly by putting $p_0=1-p_3=\eps$ with $\eps\ll
1$. In this way the empty and fully occupied configurations are no 
longer absorbing. This brings the model back into the class of
equilibrium models for which there is no phase transition in one
dimension but metastable states can nevertheless persist for long
times. The mean field equation for the density $c$ becomes
\meq[eq:c_eps]{ c'=&\eps(1-c)^3 + 3p_1c\left(1-c\right)^2+\\
&3p_2c^2\left(1-c\right) + (1-\eps)c^3. }

We study the asymptotic density as $p_1$ and $p_2$ move on a line
with slope 1 inside the hatched region of Fig.~\ref{fig:meanfield}.
For $p_1$ close to zero, Eq.~\eqref{eq:c_eps} has only one fixed
point, which is stable and close to $\eps$. As $p_1$ increases
adiabatically (by taking $c$ at $t=0$ equal the previous value of
the fixed point) the new asymptotic density will still assume this
value even when two more fixed points appear one of which is unstable
and the other stable and close to one. Eventually the first fixed
point disappears, and the asymptotic density jumps to the stable
fixed point close to one. Going backwards on the same line, the
asymptotic density will be close to one until that fixed point
disappears and it will jump back to a small value close to zero. By
symmetry, the hysteresis loop is centered around the line $p_1+p_2=1$
which we identify as a first-order phase transition line inside the
hatched region.

\begin{figure}[t]
 \centerline{\psfig{figure=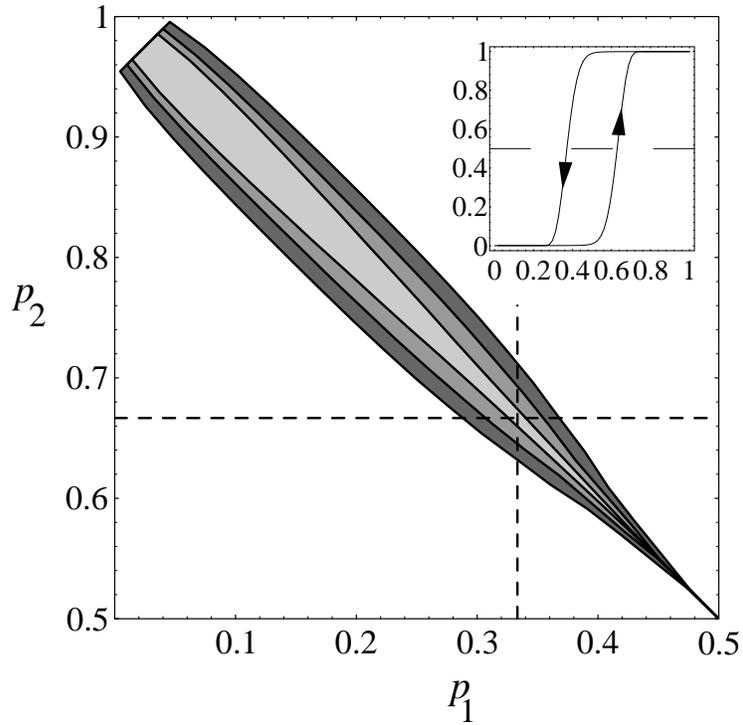,width=10cm}}
 \caption{
 Profile of the hysteresis region for several values of the noise $\eps$
 and relaxation time $T$ according to the local structure approximation
 with $l=6$. The curves represent the intersections of the hysteresis
 cycle with $c=0.5$ (horizontal dashed line in the inset). The curves
 join smoothly at $p_1=0$, $p_2=1$ (not represented). Starting from the
 out most curve, these correspond to $T=500$, $\eps=0.0001$; $T=1000$,
 $\eps=0.0001$, and $T=500$, $\eps=0.001$. The dashed lines represent
 the mean-field hysteresis region. The inset represents the cycle along
 a line parallel to the diagonal $p_1=p_2$. 
 }
 \label{fig:hysteresis}
\end{figure}

The hysteresis region is found by two methods, the dynamical mean
field, which extends the mean field approximation to blocks of $l$
sites~\cite{Gutowitz_etal_1987}, and direct numerical experiments.
As stated before it is necessary to introduce a small perturbation 
$\eps=p_0=1-p_3$. We consider lines parallel to the diagonal
$p_1=p_2$ in the parameter space and increase the value of $p_1$ and
$p_2$ after a given relaxation time $t_r$ up to $p_2=1$; afterwards
the scanning is reverted down to $p_1=0$. The hysteresis region for
various values of parameters are reported 
Fig.~\ref{fig:hysteresis}. In numerical simulations, one can
estimate the size of the hysteresis region by starting with
configurations $\vec{0}$ and $\vec{1}$, and measuring the size $d$ of
the region in which the two simulations disagree.

\section{Reconstruction of the potential}\label{sec:reconstruction}

\begin{figure}[t]
 \centerline{\psfig{figure=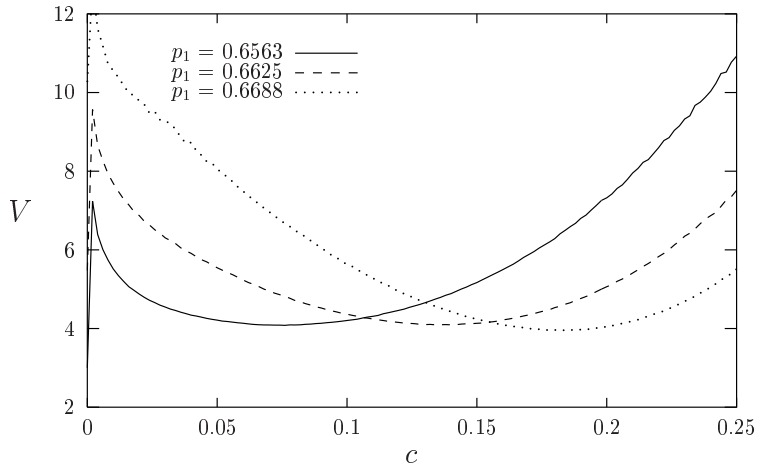,width=10cm}}
 \caption{Reconstruction of potential $V(c)$ for 
 $p_2=0$. We performed $10^4$ runs over a system of 500 sites and computed the probability
distribution $P(n)$ averaging over 500 time steps after discarding 1000 time
steps. We limited to small times in order to be able to see the divergence
at the origin (the absorbing state) together with the other local minimum
(the active state). 
 }
 \label{fig:3inp}
\end{figure}

\begin{figure}[t]
 \centerline{\psfig{figure=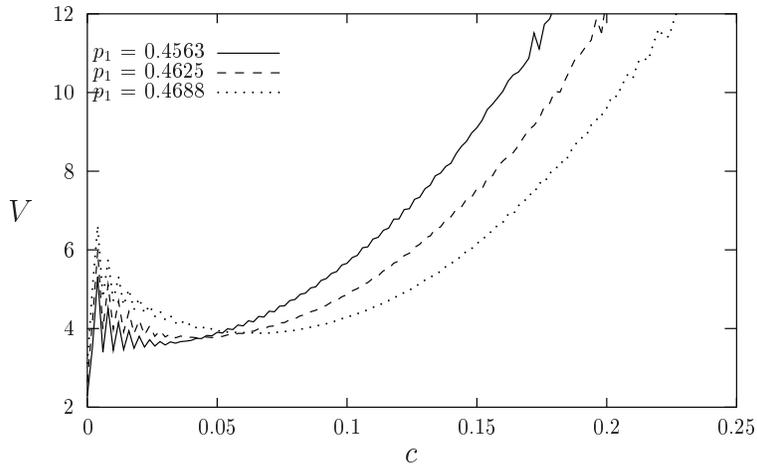,width=10cm}}
 \caption{Reconstruction of potential $V(c)$ for the kink dynamics 
 on the line $p_2=1-p_1$. The simulations were carried out as described in the
 previous figure caption.}
 \label{fig:a}
\end{figure}

An important point in the study of systems exhibiting absorbing states is 
the
formulation of a coarse-grained description using a Langevin equation.
It is generally accepted that DP universal behavior is represented by 
\eq[laplacian]{
 \dfrac{\de c(x,t)}{\de t} =  
 a c(x,t) -b c^2(x,t) + \nabla^2 c(x,t) +
 \sqrt{c(x,t)}\alpha(x,t),
}
where $c$ is the density field, $a$ and $b$ are control parameters
and $\alpha$ is a Gaussian noise with correlations
$\langle \alpha(x,t) \alpha(x',t')\rangle =
\delta_{x,x'}\delta_{t,t'}$. The diffusion coefficient has been
absorbed into the parameters $a$ and $b$ and the time scale. This
equation can be obtained by a mean-field approximation of the
evolution equation keeping only the relevant terms. The state
$c(x,t)= 0$ is clearly stationary, but its absorbing character is
given by the balance between fluctuations, which are of order of the
field itself, and the ``potential'' part  $a c -b c^2$ (See also
Ref.~\cite{Munoz_1998}).

The role of the absorbing state can be illustrated by taking a
sequence of equilibrium models whose energy landscape exhibits
the coexistence of an infinitely deep well (the absorbing state) 
and another broad local minimum 
(corresponding to the ``active'',
disordered state), separated by an energy barrier. 
There is no true
stationary active state for such a system (with a finite energy barrier), 
since there is always a
probability of jumping into the absorbing state. 
However, the system can survive 
in a metastable active state for time intervals of the order of the inverse 
of the heigth of the energy barrier. 
The parameters controlling the height of the energy
barrier are the size of the lattice and the length of the simulation:
the equilibrium systems are two-dimensional with asymmetric
interactions in the time direction~\cite{Georges_1989}.
In the limiting
case of an infinite system, 
the height of the energy barrier is finite below the transition
point, and infinite above, when the only physically relevant state is the
disordered one.  

It is possible to introduce a zero-dimensional approximation to the model by
averaging over the time and the space, assuming
that the system has entered the metastable state.  In this
approximation, the size of the original systems enters through the
renormalized coefficients  $\overline{a}$, $\overline{b}$,  
\eq{
 \dfrac{\de c(x,t)}{\de t} =  
 \overline{a} c(x,t) -\overline{b} c^2(x,t) +
 \sqrt{c(x,t)}\alpha(x,t),
}
where also the time scale has been renormalized.

The associated Fokker-Planck equation is
\eq{
 \dfrac{\de P(c,t)}{\de t} = - \dfrac{\de}{\de c}(\overline{a}c
 -\overline{b}c^2)P(c,t)
 + \dfrac{1}{2} \dfrac{\de^2}{\de c^2}c P(c,t),
}
where $P(c,t)$ is the probability of observing a density $c$ at
time $t$. One possible solution is a $\delta$-peak centered at the origin, 
corresponding to the absorbing state. 

By considering only those trajectories
that do not enter the absorbing state during the observation time, one
can impose a detailed balance condition, whose effective agreement
with the actual probability distribution has to be checked \emph{a
posteriori}. 

A stationary distribution $P(c)=\exp(-V(c))$ 
corresponds to an effective potential $V(c)$ of the form
\eq{
 V(c) = \log(c) - 2\overline{a}c + \overline{b} c^2.
}
Note that this distribution is not normalizable. One can 
impose a cutoff for low $c$, making $P(c)$ normalizable.
For finite systems the only stationary solution is the absorbing state.
However, by increasing the size of the system, one approximates the
limit in which the energy barrier is infinitely hight, the
absorbing state unreachable and $P(c)$ is the observable distribution. 

In order to find the form of the effective potential for spatially
extended systems, Mu\~noz~\cite{Munoz_1998} numerically integrated
Eq.~\eqref{laplacian}, using the procedure described by
Dickman~\cite{Dickman_1994}. It is however possible to obtain the
shape of the effective potential from the actual simulations, simply
by plotting $V(c) = -\log(P(c))$ versus $c$, where $c$ is the density
of the configuration.

In Fig.~\ref{fig:3inp} we show the profile of the reconstructed
potential $V$ for some values of $p$ around the critical value
or the infinite system
$p^*_1$ on the line $q=0$.
We used rather small systems and followed the evolution for a
limited amount of time in order to balance the weight of the
 $\delta$-peak with respect to $P(c)$ (which is only metastable).
For larger systems the absorbing
state is not visible above the transition and dominates below it. 

On the line $q=0$ the model belongs to the DP
universality class. One can observe that the curve becomes broader in
the vicinity of the critical point, in correspondence of the
divergence of critical fluctuations $\chi \sim |p-p_c|^{-\gamma'}$,
$\gamma'=0.54$~\cite{Munoz_etal_1999}. By repeating the same type of
simulations for the kink dynamics (random initial condition), we
obtain slightly different curves, as shown in Fig.~\ref{fig:a}. We
notice that all curves have roughly the same width. Indeed, the
exponent $\gamma'$ for systems in the PC universality class is
believed to be exactly 0~\cite{Jensen_1994}, as given by the scaling
relation~\cite{Munoz_etal_1999} $\gamma' = d \nu_\perp - 2\beta$.
Clearly, much more informations can be obtained from the knowledge
of $P(c)$, either by direct numerical simulations or dynamical mean
field trough finite scale analysis, as shown for instance in
Ref.~\cite{Jensen_Dickman_1993}.

\section{Discussions and conclusions}

We have studied a probabilistic cellular automata with two absorbing
states and two control parameters. This is a simple and natural
extension of the Domany-Kinzel (DK) model. Despite its simplicity it
has a rich phase diagram with two symmetric second-order phase curves
that join a first-order line at a bicritical point. The phase diagram
and the critical properties of the model were found using several
mean field approximations and numerical simulations. The second-order
phase transitions belong to the directed percolation universality
class except for the bicritical point, which belongs to the parity
conservation (PC) universality class. The first-order phase
transition line was put in evidence by a modification of the model
that allows one to find the hysteresis cycles. The model also
presents a chaotic phase analogous to the one present in the DK
model. This phase was studied using direct numerical simulations and
dynamical mean field.

On the line of symmetry of the model the relevant behavior is given
by kink dynamics. We found a closed expression for the kink evolution
rule and studied its critical properties, which belong to the PC
universality class. The effective potential governing the
coarse-grained evolution for the DP and the PC phase was found
through direct simulations, confirming that critical fluctuations
diverge at most logarithmically in the PC class.

The phase diagram of our model is qualitatively similar to 
Bassler and Browne's (BB) one~\cite{Bassler_Browne_1996}. In both models
two critical lines in the DP universality class 
meet at a bicritical point in the PC universality class, 
and give origin to a first-order transition line. This suggests that
the observed behavior has a certain degree of universality. 

An interesting feature of the BB model is that the absorbing states
at the bicritical point are indeed symmetric, but the model does not
show any conserved quantities.  We have shown that the bicritical
dynamics of our model can be exactly formulated either in terms of
symmetric states or of kinks dynamics,
providing an exact  correspondence between the presence of conserved
quantities and the symmetry of absorbing states.

Furthermore, in order to obtain a qualitatively correct mean-field 
phase diagram  of the BB model, one has to include correlations between
triplets, while the mean-field phase diagram of our model is already
correct at first approximation. This suggests that we have described a
simpler model, which can be used as prototype for multi-critical systems.

\section*{Acknowledgements} 
Helpful and fruitful discussions with Paolo Palmerini, Antonio Politi
and Hern\'an Larralde are acknowledged. This
work benefitted from partial economic support from CNR (Italy), 
CONACYT (Mexico), project IN-116198 DGAPA-UNAM (Mexico) and project
G0044-E CONACYT (Mexico). We wish to thank one of the Referees for
having indicated us Ref.~\cite{Bassler_Browne_1996}.

\appendix
\setcounter{equation}{0}
\renewcommand{\theequation}{\thesection.\arabic{equation}}
\section{Damage spreading in the mean field approximation} 
\label{app:meandamage}

The minimum damage spreading occurs when the two replicas 
$\vec{x}$ and $\vec{y}$
evolve using maximally correlated random numbers, i.e.\ when all 
$r_s$ in Eq.~\eqref{x} are the same.~\cite{Domany_Hinrichsen}. 
Let $w= x\XOR y$ be the damage at a site $i$ and time $t$. 
It is also possible to 
consider $w$ as an independent variable and write $y=x\XOR w$.
We denote $s=x_-+x+x_+$,
$s'=y_-+y+y+=(x_-\XOR w_-)+(x\XOR w)+(x_+\XOR w_+)$
and $s''=w_- + w + w_+$. The evolution equation for 
$h$, the density of damaged sites $w$ at time $t$, is obtained by considering 
all the local configurations $x_- x x_+$ and $w_- w w_+$ of one replica
and of the damage 
\eq[eq:eta]{
 h' = 
 \sum_{\substack{{x_-xx_+}\\{w_-ww_+}}} 
 \pi(c,s,3)\pi(h,s'',3) \big| p_s-p_{s'} \big|,
}
where 
\eq{
 \pi(\alpha, n,m) = \alpha^{n}(1-\alpha)^{n-m}.
}
In this 
equation all the sums run from zero to one.
The value of $c$ is given by Eq.~\eqref{eq:c}. The term $| p_s-p_{s'}|$
is the probability that $R_s\XOR R_{s'}$ is 
one using only one random number for the $r_s$. The argument of the sum is the
probability that $x'\neq y'$.
It is possible to rewrite Eq.\eqref{eq:eta} in a different form
\eq[eq:damage1]{
 h'=\sum_{ss'\ell}\binom{m}{s}\binom{m-s}{\ell}\binom{m-s}{s'-\ell}%
 \pi(c,s,m)\pi(\eta,s+s'-2\ell,m)\lvert p_s-p_{s'}\rvert,
}
where $s$ and $s'$ are the same as above, and $\ell$ is the overlap between $x$
and $y$, i.e.\ 
$\ell = (x_-\AND y_-)+(x\AND y)+(x_+\AND y_+)$
($\AND$ is the AND operation). Assuming that $\binom{a}{b}=0$ 
if $b>a$, $a<0$ or 
$b<0$, the sum of \eqref{eq:damage1} can run over all positive integers.
This expression is valid for all totalistic 
rules with a neighborhood of size $m$
 (here $m=3$).
 
The stationary 
state of Eq.~\eqref{eq:eta} (or Eq.~\eqref{eq:damage1})
can be found 
analytically using a symbolic manipulation program.
The chaotic
transition line is
\eq{
 p_2 = p_1-\dfrac{1}{9},
}
with $1/3<p_1<1$,$0<p_2<2/3$.

\section{Kink dynamics}
\label{app:kinks}

On the segment $p_1+p_2=1$, $p_1<p_t$ the order parameter is the number 
of kinks. The dynamics of the 
kinks $y_i=x_i\XOR x_{i+1}$ (that for the ease of notation we write
$y=x\XOR x_+$) is obtained by taking the exclusive disjunction of 
$x'=x_i^{t+1}$ and $x'_+=x_{i+1}^{t+1}$
given by Eq.~\eqref{x}. In order 
to obtain a closed expression for the $y$, a little of Boolean algebra is 
needed. 

The totalistic functions $\delta_{\sigma,s}$ 
 where $s=x_-+x+x_+$ can be expressed in terms of the 
symmetric polynomials $\xi^{(j)}$ of degree $j$~\cite{Bagnoli_1992}. 
 These are
\meq{
 \xi^{(1)} &= x_-\XOR x\XOR x_+,\\
 \xi^{(2)} &= x_-x \XOR x_-x_+ \XOR xx_+,\\
 \xi^{(3)} &= \xi^{(1)} \xi^{(2)} = x_-xx_+.
} 
The totalistic functions are given by
\meq{
 \delta_{\sigma,1} &= \xi^{(1)} \XOR \xi^{(3)},\\
 \delta_{\sigma,2} &= \xi^{(2)} \XOR \xi^{(3)},\\
 \delta_{\sigma,3} &= \xi^{(3)}.
}

In the evolution equation \eqref{x}, one 
has $R_1 = 1$ if $r_1<p_1$, and $R_2= 1$ if $r_2<p_2$.
On the line $p_1+p_2=1$ (i.e.\ $p_2=1-p_1$), 
$R_2$ takes the value 1 if $1-r_2>p_1$. Choosing 
$1-r_2=r_1$ (this choice does not affect the dynamics of a single
replica) we have $R_2 = R_1\XOR 1$
and Eq.~\eqref{x} becomes, after some manipulations,
\eq{
 x' = R (\xi^{(1)} \XOR \xi^{(2)}) \XOR \xi^{(2)},
}
where 
$R= R_1$.
One can easily check that 
\eq{ 
 \xi^{(1)} = y_- \XOR y \XOR x,
}
and 
\eq{
 \xi^{(2)} = y_-y \XOR x.
} 
Finally, we obtain the evolution equation for the $y$
\meq[y]{
 y' &= x' \XOR x'_+ \\ 
 &=R (y_-\XOR y \XOR y_-y) \XOR R_+(y\XOR y_+ 
 \XOR yy_+) \XOR y_-y \XOR yy_+ \XOR y \\
 &= R (y_-\OR y) \XOR R_+ (y \OR y_+) 
 \XOR y_-y \XOR yy_+ \XOR y,
}
In this expression $\OR$ denotes the disjunction operation (OR). 
The sum modulo 2 (XOR) of all $y_i$ over the lattice is invariant with time,
since all repeated terms cancel out ($a \XOR a = 0$). 
Note that the kink dynamics uses
correlated noise between neighboring sites.

\end{document}